\documentclass[prx,twocolumn,floatfix,superscriptaddress]{revtex4}
\usepackage{graphics,epsfig,amsfonts,amssymb,amsmath}
\usepackage{soul} 
\usepackage[dvipsnames]{xcolor}
\usepackage{MnSymbol,wasysym}
\usepackage{bm}

\newcommand{\beginsupplement}{%
\onecolumngrid

\pagebreak

\begin{center}
  \textbf{\large Supplementary Material}\\[.2cm]
\end{center}

				\setcounter{section}{0}
        \renewcommand{\thesection}{S\arabic{section}}%
        \setcounter{table}{0}
        \renewcommand{\thetable}{S\arabic{table}}%
        \setcounter{figure}{0}
        \renewcommand{\thefigure}{S\arabic{figure}}%
				\setcounter{equation}{0}
				\renewcommand{\theequation}{S\arabic{equation}}
				
     }

\newcommand{\beginmethods}{%

\vfill

  \textbf{\large Methods}

}

\newcommand{\beginacknowd}{%

\vfill

  \textbf{\large Acknowledgements}

}

\newcommand{\begindatav}{%

\vfill

  \textbf{\large Data Availability}

}

\newcommand{\beginauthorcont}{%

\vfill

  \textbf{\large Author Contributions}

}

\newcommand{\begincompint}{%

\vfill

  \textbf{\large Competing Interests}

}

\newcommand{\idop}{1\!\!1}

\begin{document}
\title{Deconfinement of Mott Localized Electrons into Topological and Spin-Orbit Coupled Dirac Fermions}

\author{Jos\'e M. Pizarro}
\email{jpizarro@uni-bremen.de}
\affiliation{Institute for Theoretical Physics, University of Bremen, Otto-Hahn-Allee 1, 28359 Bremen, Germany}
\affiliation{Bremen Center for Computational Material Sciences, University of Bremen, Am Fallturm 1a, 28359 Bremen, Germany}
\author{Severino Adler}
\affiliation{Institut f\"ur Theoretische Physik und Astrophysik and W\"urzburg-Dresden Cluster of Excellence ct.qmat, Universit\"at W\"urzburg, 97074 W\"urzburg, Germany} 
\affiliation{Institute of Solid State Physics, TU Wien, A-1040 Vienna, Austria}
\author{Karim Zantout}
\affiliation{Institute of Theoretical Physics, Goethe University Frankfurt am Main, D-60438 Frankfurt am Main, Germany}
\author{Thomas Mertz}
\affiliation{Institute of Theoretical Physics, Goethe University Frankfurt am Main, D-60438 Frankfurt am Main, Germany}
\author{Paolo Barone}
\affiliation{Consiglio Nazionale delle Ricerche, Institute for Superconducting and Innovative Materials and devices (CNR-SPIN), c/o University "G. D'Annunzio" , Chieti 66100, Italy}
\author{Roser Valent\'i}
\affiliation{Institute of Theoretical Physics, Goethe University Frankfurt am Main, D-60438 Frankfurt am Main, Germany}
\author{Giorgio Sangiovanni}
\affiliation{Institut f\"ur Theoretische Physik und Astrophysik and W\"urzburg-Dresden Cluster of Excellence ct.qmat, Universit\"at W\"urzburg, 97074 W\"urzburg, Germany}
\author{Tim O. Wehling}
\email{twehling@uni-bremen.de}
\affiliation{Institute for Theoretical Physics, University of Bremen, Otto-Hahn-Allee 1, 28359 Bremen, Germany}
\affiliation{Bremen Center for Computational Material Sciences, University of Bremen, Am Fallturm 1a, 28359 Bremen, Germany}


%
\date{\today}
\begin{abstract}  
The interplay of electronic correlations, spin-orbit coupling and topology holds promise for the realization of exotic states of quantum matter. Models of strongly interacting electrons on honeycomb lattices have revealed rich phase diagrams featuring unconventional quantum states including chiral superconductivity and correlated quantum spin Hall insulators intertwining with complex magnetic order. Material realizations of these electronic states are however scarce or inexistent. In this work, we propose and show that stacking  1T-TaSe$_2$ into bilayers can deconfine electrons from a deep Mott insulating state in the monolayer to a system of correlated Dirac fermions subject to sizable spin-orbit coupling in the bilayer. 1T-TaSe$_2$ develops a Star-of-David charge density wave pattern in each layer. When the Star-of-David centers belonging to two adyacent layers are stacked in a honeycomb pattern, the system realizes a generalized Kane-Mele-Hubbard model in a regime where Dirac semimetallic states are subject to significant Mott-Hubbard interactions and spin-orbit coupling. At charge neutrality, the system is close to a quantum phase transition between a quantum spin Hall and an antiferromagnetic insulator. We identify a perpendicular electric field and the twisting angle as two knobs to control topology and spin-orbit coupling in the system. Their combination can drive it across hitherto unexplored grounds of correlated electron physics including a quantum tricritical point and an exotic first-order topological phase transition.
\end{abstract}


\maketitle

\section{Introduction}
Prospects of quantum information technologies have motivated an intense search for systems which intertwine topology and electronic correlations \cite{FQHE_82,qi_topological_2011,sato_topo_SC_review_2017,Rachel_2018}. Strongly interacting and spin-orbit coupled electrons on the honeycomb lattice, as theoretically described by the Kane-Mele-Hubbard model \cite{hohenadler_correlation_2013}, feature quantum spin Hall (QSH), Mott-Hubbard, collective magnetic and chiral superconducting states. While several weakly interacting QSH systems are now well-studied experimentally \cite{qi_topological_2011,wehling_dirac_2014}, material realizations of honeycomb Kane-Mele-Hubbard fermions with strong correlations and spin-orbit coupling are rare \cite{marrazzo_prediction_2018, wu_unconventional_2019}. Here, we introduce a new 'van der Waals engineering' platform to serve this purpose.

\begin{figure*}[h]
\includegraphics[width=0.85\textwidth]{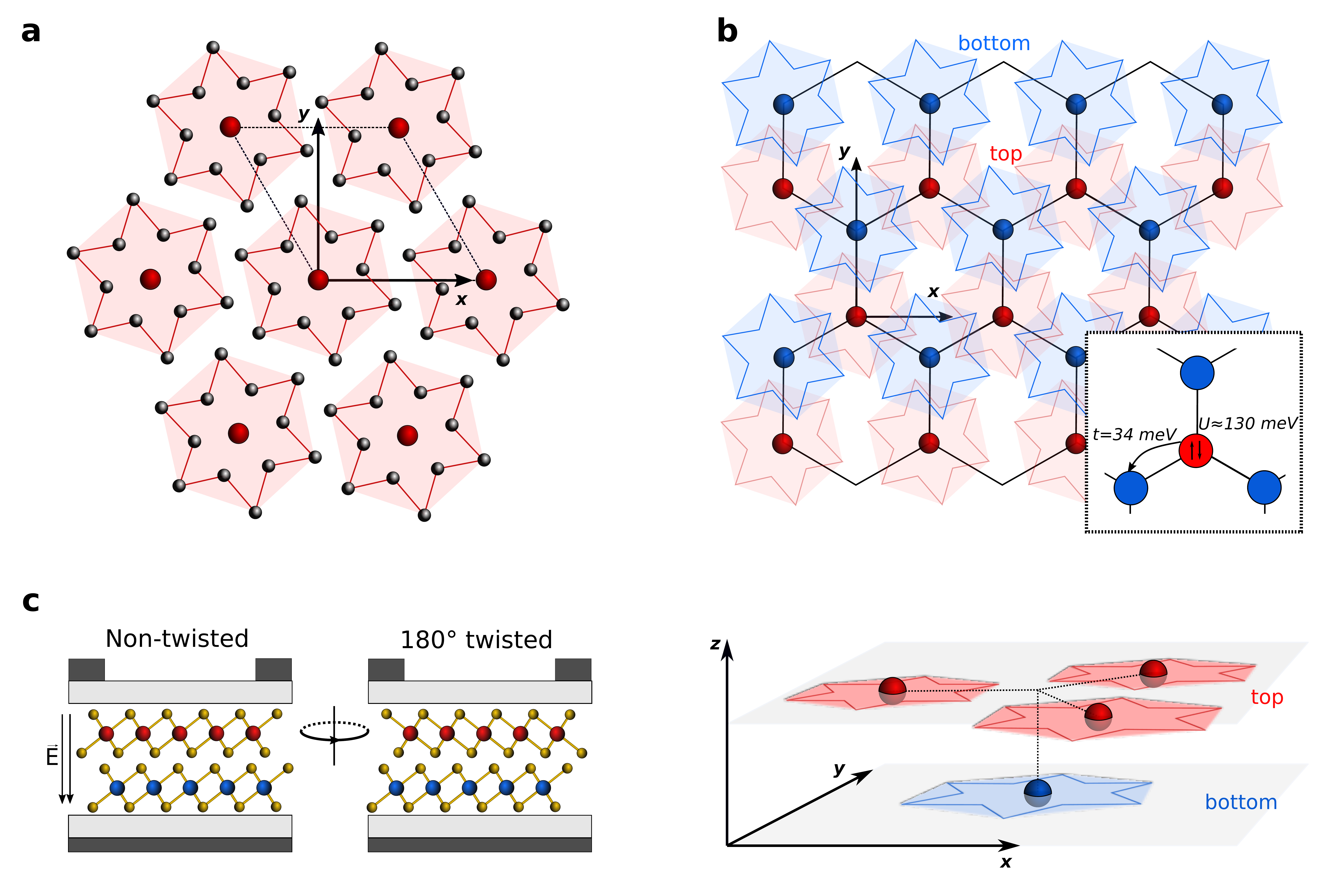} 
\caption{\textbf{Crystal structures of 1T-TaSe$_2$ mono- and bilayers in the CCDW phase.} \textbf{a}, Monolayer 1T-TaSe$_2$ in the CCDW phase. Only Ta atoms are shown. The Ta atoms are distorted into a SoD pattern, where the central Ta atoms (red) are surrounded by two rings of in total twelve Ta atoms (black). The SoDs are marked with red lines as guide to the eyes. \textbf{b}, Top and three-dimensional side view of honeycomb stacked CCDW 1T-TaSe$_2$ bilayer. Only the Ta atoms in the SoD centers are shown, with blue spheres and shaded regions marking the bottom layer atoms, and red spheres and shaded regions for the top layer. 
The inset illustrates the leading terms of the effective low-energy Hamiltonian, i.e. the nearest-neighbor hopping $t=-34$~meV and the local interaction $U\approx 130$~meV. \textbf{c}, Side view of 1T-TaSe$_2$ bilayers embedded in field effect transistor structures for the application of vertical electric fields. A non-twisted and a 180$^\circ$ twisted bilayer are shown with Ta atoms (red and blue) and Se atoms (yellow). 
} 
\label{fig1:structures}%
\end{figure*}

We show that stacking of 1T-TaSe$_2$ into bilayers can deconfine electrons from a deep Mott insulating state realized in the monolayer to a system of correlated Dirac fermions subject to sizable spin-orbit coupling. Central to this transition is the possibility of van der Waals materials to stack in different configurations. For a specific honeycomb arrangement (Fig. \ref{fig1:structures}), the kinetic energy associated with the electronic hopping $t$ turns out to be of the same order of magnitude as the effective local Coulomb repulsion $U$. The system features therefore electronic correlations, which turn out to put the system right on the verge between QSH and correlated antiferromagnetic insulating states at charge neutrality and support chiral superconductivity under doping. We finally demonstrate that tuning the system via electric fields and twisting of the layers relative to each other sensitively affects the low-energy electronic structure in terms of emerging Dirac mass and spin-orbit coupling terms and leads to completely unexplored regimes of correlated electrons. 


\section{Results and Discussion}
\subsection{From a correlated insulator to emergent Dirac fermions}
Layered group-V transition metal dichalcogenides (TMDCs) such as 1T-TaSe$_2$ or 1T-TaS$_2$ feature a low-temperature commensurate charge density wave (CCDW) where Ta-atoms are displaced into Star-of-David (SoD) patterns (Fig. \ref{fig1:structures}a) \cite{wilson_charge-density_1975, rossnagel_origin_2011}. In this phase, the SoDs form a triangular $\sqrt{13}\times\sqrt{13}$ superlattice in every layer and host correlated electrons: 1T-TaS$_2$ shows a metal-to-insulator transition when entering the CCDW phase \cite{rossnagel_origin_2011}. In 1T-TaSe$_2$, the bulk remains conductive till lowest temperatures, while the surface exhibits a Mott transition around 250~K \cite{perfetti_spectroscopic_2003,colonna_mott_2005}. Recently, 1T-TaSe$_2$ has been fabricated down to monolayer thickness \cite{nakata_selective_2018,borner_observation_2018,chen2019visualizing} and a pronounced thickness dependence of the electronic structure has been reported \cite{chen2019visualizing}. As bonds between the layers are mainly of van der Waals type, different stacking configurations are observed in experiments \cite{Rosenauer_TEM_2018, Kourkoutis_TEM_2016} and have a strong impact on the electronic structure \cite{freericks_pruschke_2009,marianetti_2014,ritschel_stacking-driven_2018}.

We compare the CCDW state in the monolayer to a bilayer with honeycomb stacking, where the SoD centers form a buckled honeycomb lattice (Fig. \ref{fig1:structures}b). The bottom layer SoD centers form sublattice A and the top layer sublattice B. This stacking is one of many possible configurations, which can generally differ, both, by the local atomic stacking and by the stacking of the SoD centers. The local stacking considered here, has the Ta atoms of the bottom layer approximately beneath the lower Se atoms of the top layer (Fig. \ref{fig1:structures}c). While this kind of local stacking is not the most commonly observed one of the T-phase TMDCs \cite{Kourkoutis_TEM_2016,chen2019visualizing}, a corresponding stacking sequence has been found in transmission electron microscopy studies of 1T-TaS$_2$ \cite{Kourkoutis_TEM_2016} and turns out to be metastable in density functional theory (DFT) simulations of 1T-TaSe$_2$ \cite{PhysRevB.92.224104}. Regarding the stacking of the SoD centers further van der Waals DFT total energy calculations (see Supplementary Information) show that the particular honeycomb arrangement studied here, is metastable and energetically on the order of 10~meV per formula unit above the lowest energy configuration. This is similar to the configuration observed experimentally in Ref. \cite{chen2019visualizing} and it is thus plausible that also the honeycomb configuration considered here, is within reach of experiments. Experimental approaches including tear-and-stack \cite{weston_atomic_2020} and STM voltage pulse based manipulation schemes \cite{ma_metallic_2016} can present possible avenues to control and switch metastable stacking configurations and to reach the honeycomb configurations discussed in our paper. At small twist angles all different kinds of configurations can be realized locally in the moiré including likely those honeycomb cases discussed here.


To study the electronic structure of such engineered stackings, we combine \textit{ab-initio} calculations in the frameworks of DFT and the random phase approximation (RPA) with effective low-energy models, which we investigate with dynamical mean-field theory (DMFT) and two-particle self-consistent (TPSC) many-body approaches, see Methods section.


\begin{figure*}
\includegraphics[width=0.85 \linewidth]{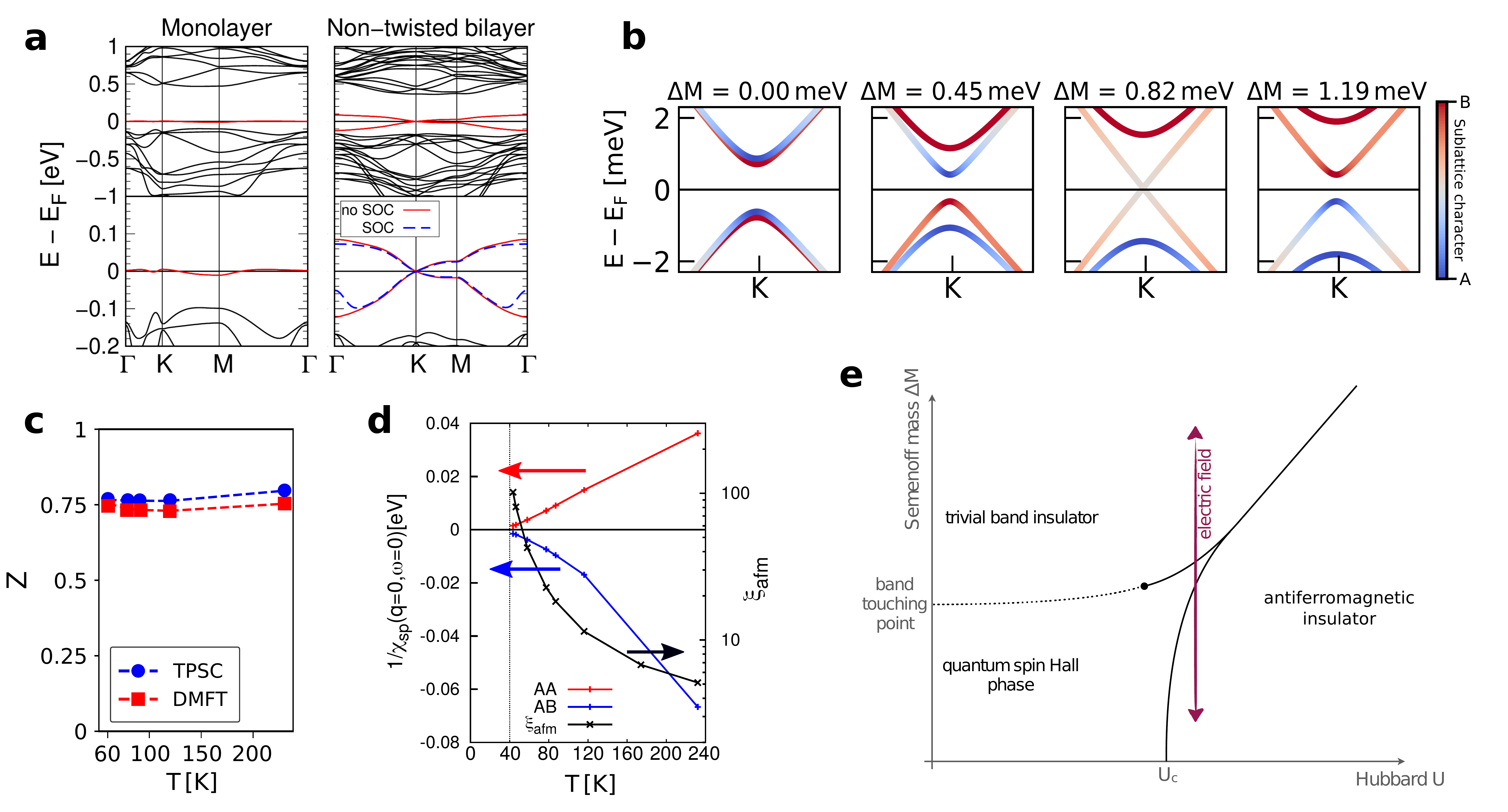}
\caption{\textbf{Electronic structure and phase diagram of CCDW 1T-TaSe$_2$ systems.} \textbf{a}, Band structures of CCDW 1T-TaSe$_2$ monolayer (left) in comparison to honeycomb stacked non-twisted CCDW 1T-TaSe$_2$ bilayer (right). Bottom panels show a zoom around the Fermi level $E_F$. The red and dashed blue lines mark the DFT low-energy bands without and with SOC included, respectively. From the flat band near $E_F$ in the monolayer (red solid line, bandwidth $\approx 14$~meV) two dispersive bands with a bandwidth $\approx 200$~meV emerge in the bilayer case. The bilayer bands exhibit Dirac points in the Brillouin zone corners, K and K'. \textbf{b}, Influence of extrinsic Semenoff mass terms $\Delta M$ on the low-energy band structure. The sublattice character is color coded. The system changes from QSH to trivial band insulator at $\Delta M=0.82$~meV, which corresponds to a vertical electric field of $E_{\rm z} \approx 0.5$~mV~\AA$^{-1}$. \textbf{c}, Quasi-particle weight $Z$ for non-twisted CCDW 1T-TaSe$_2$ bilayer calculated with DMFT and TPSC. Both approaches place the system consistently in the moderately correlated regime $Z\approx 0.75$ at all calculated temperatures. \textbf{d}, Temperature-dependent antiferromagnetic correlation length $\xi_{\rm AFM}$ and inverse static spin susceptibilities of non-twisted CCDW 1T-TaSe$_2$ bilayer at wave vector $\mathbf{q}=\mathbf{0}$ as calculated with TPSC. The intra-sublattice ($1/\chi_{AA}=1/\chi_{BB}$) and inter-sublattice ($1/\chi_{AB}$) elements of the inverse susceptibility at wave number $\mathbf{q}=\mathbf{0}$ are shown. \textbf{e}, Schematic phase diagram of honeycomb stacked non-twisted CCDW 1T-TaSe$_2$ bilayer as a function of extrinsic Semenoff mass $\Delta M$ and interaction strength $U$. The region accessible for non-twisted CCDW 1T-TaSe$_2$ bilayer through tuning with external electric fields is highlighted. The transition from QSH to band insulator is a continuous transition at small $U$ (dashed line) and a first order transition at larger $U$ (solid line). The red area in the quantum spin Hall region indicates the increasing many-body character of this phase. 
}
\label{fig2:bands} 
\vspace{-0.3cm}
\end{figure*}

In the CCDW phase, the band structure of monolayer 1T-TaSe$_2$ obtained with non-spin-polarized DFT is characterized by a single (Ta) flat band at the Fermi level \cite{rossnagel_origin_2011,freericks_pruschke_2009,marianetti_2014,ritschel_stacking-driven_2018}, which has a bandwidth of less than 20~meV (Fig. \ref{fig2:bands}a, left). Hence, the CCDW formation largely quenches in-plane hopping of the electrons. In the honeycomb stacked bilayer (with no twist), two dispersive bands with a bandwidth of the order of 200~meV emerge from the low-energy flat band of the monolayer (Fig. \ref{fig2:bands}a, right). Comparison of the mono- and bilayer bandwidths shows that interlayer hopping effects must dominate over intralayer hoppings by approximately an order of magnitude. In this sense, CCDW TaSe$_2$ bilayers are the exact opposite of graphene bilayer systems, since in the latter out-of-plane coupling is an order of magnitude weaker than in-plane hopping \cite{castro_neto_RMP_2009,katsnelson_2012}. For the  1T-TaSe$_2$ honeycomb bilayer, the upper and lower low-energy bands touch as Dirac points at the Brillouin zone corners K and K'. In the undoped system, these Dirac points are exactly at the Fermi level.


We next construct a Wannier Hamiltonian to describe the Dirac bands with one Wannier function for each SoD center, i.e. two Wannier orbitals per bilayer CCDW superlattice unit cell (Supplemental Figure 2). The resulting nearest-neighbor hopping between a sublattice A site in the bottom and a neighboring sublattice B site in the top layer amounts to $t=-34$~meV and is the leading term of the Wannier Hamiltonian. There are further terms in the Wannier Hamiltonian, which are, however, at least an order of magnitude smaller than $t$.
	
The effective Hubbard interaction $U$ for the SoD Wannier orbitals of CCDW TaSe$_2$ calculated in RPA is $U\approx130$~meV, which is also in line with the experimental estimates in Ref. \cite{chen2019visualizing} and calculations for TaS$_2$ \cite{marianetti_2014}. The ratio of hopping to Coulomb interaction is decisive in determining the strength and kind of electronic correlation phenomena taking place. Our calculations yield $U/|t|\approx 3.8$.


To study the resulting electronic correlations, we performed simulations of the Hubbard model for the non-twisted CCDW 1T-TaSe$_2$ bilayer in the framework of DMFT and the TPSC approach \cite{Zantout2018,zantout_effect_2019}. The quasi-particle weight $Z$ shown in Fig. \ref{fig2:bands}c is a measure of the electronic correlation strength. In the temperature range $T=60-230$~K, both DMFT and TPSC consistently yield essentially constant $Z\approx 0.75$. Our system is thus at intermediate local correlation strength and far away from the paramagnetic Mott-Hubbard transition taking place above $U_{\rm Mott}/t\approx 8.2$. The DMFT and TPSC studies of the full Hubbard model for the non-twisted CCDW 1T-TaSe$_2$ bilayer are, thus, in line with studies of the idealized Hubbard model on the honeycomb lattice \cite{Kuroki_DMFT_2009}. 

The TPSC calculations give insight to spin-fluctuations taking place in the system (Supplemental Figure 4). The inverse intra- and inter-sublattice terms of the static magnetic susceptibility at wave number $\mathbf{q}=\mathbf{0}$ as well as the antiferromagnetic correlation length $\xi_{\text{AFM}}$ are shown in Fig. \ref{fig2:bands}d. We observe that antiferromagnetic fluctuations with alternating spin orientation between the two sublattices (A, B) are dominant and strongly enhanced at temperatures $T\lesssim 100$~K. These fluctuations indicate that the non-twisted CCDW 1T-TaSe$_2$ bilayer is close to a quantum phase transition from a Dirac semimetal to an antiferromagnetic insulator, which occurs for ideal Hubbard honeycomb systems exactly in the range of $U_{\rm c}/t\approx 3.6-3.8$ \cite{wehling_dirac_2014,tremblay_TPSC_honeycomb_2015,HoneckerAssaad_2019}.


\subsection{Spin-orbit coupling}
The aforementioned magnetic correlation phenomena are sensitive to details of the low-energy electronic structure and spin-orbit coupling (SOC), which we discuss in the following based on a symmetry analysis. The space group of non-twisted CCDW 1T-TaSe$_2$ bilayer in the honeycomb structure is $P\bar{3}$ (\#147), comprising inversion symmetry and three-fold rotations $C_3$ around an axis perpendicular to the bilayer. Imposing also time-reversal symmetry, every band must be two-fold degenerate. Therefore, SOC-induced qualitative changes of the band structures can occur near the Dirac points at K and K'. 
A corresponding $\mathbf{k} \cdot \mathbf{p}$ expansion (see Supplementary Information) reads
\begin{equation}
\begin{aligned}
H_{\rm 0} &= \hbar v_{\rm F} (\tau k_{\rm x}S_{\rm x} + k_{\rm y}S_{\rm y}) + \lambda_{\rm SOC} \tau \sigma_{\rm z} S_{\rm z} \\
&+ \alpha_{\rm{R2}}(k_{\rm x}\sigma_{\rm y} - k_{\rm y}\sigma_{\rm x})S_{\rm z},
\end{aligned}
\label{eq:Dirac_honeycomb}
\end{equation}
where the pseudospin $\mathbf{S}$ describes the sublattice degree of freedom, $\mathbf{\sigma}$ acts on the electron spin, and $\tau=\pm 1$ labels the valley (K, K') degree of freedom. This Hamiltonian comprises three contributions; the first contribution is a two-dimensional massless Dirac term, with the sublattice-pseudospin playing the role of the spin inherent to the Dirac equation. This  term is analogous to the massless Dirac term in graphene \cite{wehling_dirac_2014,castro_neto_RMP_2009,katsnelson_2012}. 
SOC is responsible for the second and third contributions to $H_{\rm 0}$: a valley-spin-sublattice coupling $\lambda_{\rm SOC}=0.74$~meV, which is often called Kane-Mele spin-orbit term \cite{kanemele.prl95.2005,PhysRevLett.109.055502}, and a sublattice-staggered Rashba term $\alpha_{\rm{R2}}$, which belongs to the R2 class according to the classification form Ref. \cite{zhang_hidden_2014}. 
A finite Kane-Mele term $\lambda_{\rm SOC}$ opens a gap and turns the system described by $H_{\rm 0}$ into a QSH insulator \cite{kanemele.prl95.2005,PhysRevLett.109.055502}. Importantly, the Kane-Mele term here is enhanced in comparison to its counterpart in graphene by two orders of magnitude \cite{castro_neto_RMP_2009,katsnelson_2012,Kochan2017}, and corresponds to a temperature $T\approx 10$~K, which is well accessible in experiments. Given that $U/|t|\approx 3.8$, the non-twisted CCDW 1T-TaSe$_2$ bilayer implements a material realization of the Kane-Mele-Hubbard model in a regime very close to the topological quantum phase transition from a QSH insulator to an antiferromagnetic insulator (Fig. \ref{fig2:bands}e). 

Dirac fermions and their topology are affected by different kinds of mass fields. An energy difference between electrons localized in sublattice A and B leads to a so-called Semenoff mass term $M$, which would enter the Hamiltonian $H_{\rm 0}$ of equation (\ref{eq:Dirac_honeycomb}) in the form $M S_{\rm z}$. This term breaks sublattice invariance and thereby inversion symmetry and leads to a transition from a QSH to a band insulator at $|M|=|\lambda_{\rm SOC}|$ \cite{kanemele.prl95.2005, di_sante_towards_2019}.
In the non-twisted CCDW 1T-TaSe$_2$ bilayer, the intrinsic Semenoff mass $M_{\rm 0}=0$ is required to vanish by symmetry. Our Wannier construction yields $M_{\rm 0}=0.14$~meV, which is small as compared to all other relevant terms in the system. Possible origins of this small symmetry breaking can be the Wannier constructions and also faint asymmetries accumulated during self-consistency iterations in the DFT calculations. Vertical electric fields $E_{\rm z}$, as realizable in field effect transistor geometries (Fig. \ref{fig1:structures}c), break inversion symmetry, translate into staggered sublattice potentials, and therefore corresponding extrinsic Semenoff mass contributions $\Delta M$ (Fig. \ref{fig2:bands}b). DFT calculations (Supplementary Figure 3) yield the approximate relation $\Delta M\approx e E_{\rm z} d/\epsilon_{\rm \perp}$, where $d=6.4$~\AA{} is the interlayer distance, $e$ is the elementary charge, and $\epsilon_{\rm \perp}=3.72$ plays the role of an effective dielectric constant. The QSH to band insulator transition is reached for $E_{\rm z} \approx 0.5$~mV~\AA$^{-1}$$=50$~kV~cm$^{-1}$ (Fig. \ref{fig2:bands}b), which is well within reach of experiments \cite{klein_electric-field_2017}.

Taken together, our calculations show that the honeycomb-stacked non-twisted CCDW 1T-TaSe$_2$ bilayer is located in a region of the phase diagram (Fig. \ref{fig2:bands}e) with three different phases (QSH insulator, band insulator and antiferromagnetic insulator) coming together. Electron correlation is known to change the order of the QSH-band insulator transition from second to first order \cite{Griogio_first-order_2015}. Contrary to the standard non-interacting QSH to band insulator transition, where the gap closes and reopens continuously with vanishing gap at the transition point, the QSH gap remains finite and the system changes discontinuously to a band insulating state at the transition point. Application of vertical electric fields in the system at hand represents hence a possibility to realize this exotic interaction-induced first order transition.

\subsection{Twisted CCDW 1T-$\mathbf{TaSe}_2$ bilayers}

\begin{figure*}%
\includegraphics[width=0.85 \linewidth]{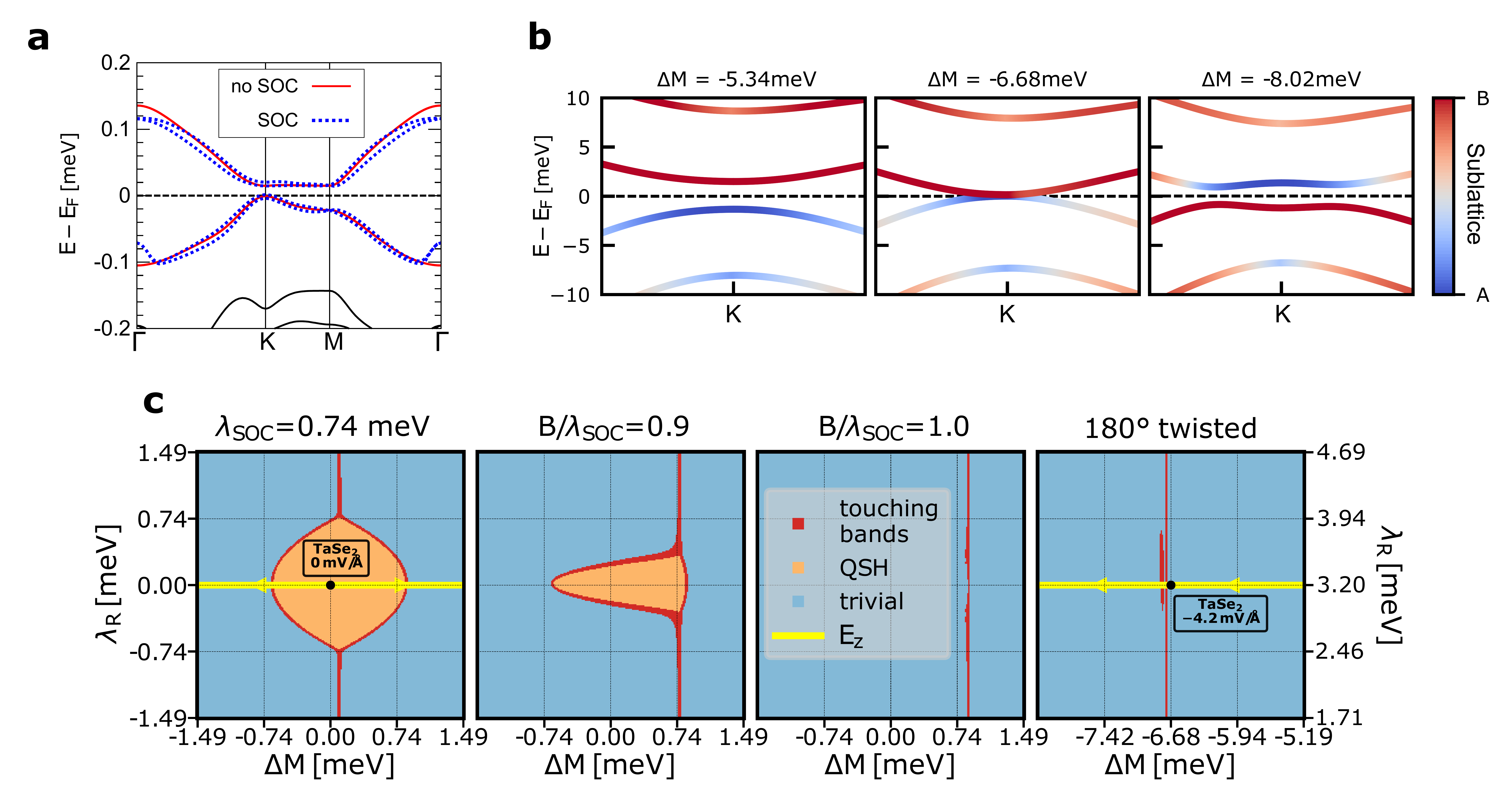}
\caption{\textbf{Influence of $\mathbf{180^{\circ}}$ twisting and electric fields on band structure and topology.} \textbf{a}, Band structure for $180^{\circ}$ twisted CCDW 1T-TaSe$_2$ bilayer without (solid red) and with (dashed blue) SOC included. The degeneracy of the bands is lifted due to the absence of the inversion symmetry after twisting. \textbf{b}, Influence of vertical electric fields on the low-energy band structure for $180^{\circ}$ twisted CCDW 1T-TaSe$_2$ bilayer. The gap at K closes and a band touching point occurs when the Semenoff mass term is $|M|=|M_{\rm 0}+\Delta M| = 1.87$~meV at $E_{\rm z}=-4.2$~mV~\AA$^{-1}$. \textbf{c}, Non-interacting $\lambda_{\rm R}$ vs $\Delta M$ topological phase diagrams for non-twisted, $180^{\circ}$ twisted CCDW bilayer and two cases in between, where the spin valley coupling $B$ of the non-twisted CCDW 1T-TaSe$_2$ bilayer is varied. For $B=0$ the system shows the symmetric onion-like shape similar to Ref. \cite{kanemele.prl95.2005}. When $B$ increases, the QSH region shrinks and disappears when $B=\lambda_{\rm SOC}$. For $B>\lambda_{\rm SOC}$, the system behaves as a trivial band insulator. The yellow arrows in the leftmost and rightmost panels indicate the path in phase space accessible by varying the vertical electric field $E_{\rm z}$.
}%
\label{fig3:manipulating}%
\end{figure*}


Layered van der Waals systems allow to realize different stacking configurations via twisting, i.e. relative rotations between the layers. General twist angles $\theta$ lead to incommensurate moir\'e patterns superimposed to the CCDW lattice.  Arguably the simplest case of twisting is a rotation angle of $\theta=180^\circ$, which leads to a system with identical Bravais lattice but different symmetry of the supercell basis (Fig. \ref{fig1:structures}c): in case of honeycomb stacking of the CCDW, the space group of the $180^\circ$ twisted structure of CCDW TaSe$_2$ bilayer is reduced to P3, meaning that inversion symmetry is lost with respect to the non-twisted case. The resulting band structure (Fig. \ref{fig3:manipulating}a) is qualitatively similar to the non-twisted honeycomb case regarding the overall shape and width of the low-energy bands. Thus, also the $180^\circ$ twisted case will be far away from the paramagnetic Mott transition. However, the low-energy band structure is markedly different in the $180^\circ$ twisted case. First, the conduction band is almost flat between $K$ and $M$. Second, inversion symmetry breaking lifts band degeneracies: our DFT calculations reveal a staggered potential and an associated intrinsic Semenoff mass term of $M_{\rm 0}=8.55$~meV, which opens a gap at the K and K' points already without SOC and without external electric fields. Furthermore, additional SOC terms are now allowed by symmetry (see Supplementary Information) and completely lift the remaining band degeneracies except for the time-reversal symmetric points $\Gamma$ and M. 

Based on a symmetry analysis, we obtain the following low-energy model in the vicinity of K and K': 
\begin{equation}
\begin{aligned}
H_{\rm{180^\circ}} &= H_{\rm 0} + M_{\rm 0}  S_{\rm z}+B \tau \sigma_{\rm z} \\
 &+\lambda_{\rm R} (\tau\sigma_{\rm y} S_{\rm x}-\sigma_{\rm x} S_{\rm y})+\alpha_{\rm{R1}} (k_{\rm x}\sigma_{\rm y}-k_{\rm y}\sigma_{\rm x})\\
&+\lambda_{\rm D} (\tau S_{\rm y} k_{\rm x}-S_{\rm x} k_{\rm y})\sigma_{\rm z}.
\end{aligned}
\label{eq:H180}
\end{equation}
The additional terms with respect to the non-twisted honeycomb structure  equation (\ref{eq:Dirac_honeycomb}) are the intrinsic Semenoff mass ($M_{\rm 0}$), the spin-valley coupling term ($B$), the Kane-Mele-Rashba interaction ($\lambda_{\rm R}$), and a Rashba interaction belonging to the R1 class \cite{zhang_hidden_2014} giving rise to pure Rashba spin-polarization patterns ($\alpha_{\rm{R1}}$). The last term (with coupling constant $\lambda_{\rm D}$) can also be seen as an effective $k-$dependent magnetic field parallel to the $z$ axis. Our DFT calculations yield $M_{\rm 0}=8.55$~meV, $B=-1.85$~meV, $|\lambda_{\rm SOC}|\lesssim 0.05$~meV and $\lambda_{\rm R}=3.21$~meV. These terms affect the dispersion and imprint an intricate sublattice and spin structure to the low-energy bands, which can be manipulated by vertical electric fields as shown in Fig. \ref{fig3:manipulating}b. 

Except for $E_{\rm z}=-4.2$~meV~\AA$^{-1}$ ($\Delta M=-6.68$meV and $M=M_{\rm 0}+\Delta M=1.87$meV), the system is always gapped. We calculated the $Z_2$ topological invariant for the non-interacting 180$^\circ$ twisted bilayer in comparison to the non-twisted bilayer case as well as for two cases in between where the ratio $B/\lambda_{\rm SOC}$ is varied (Fig. \ref{fig3:manipulating}c, and Supplemental Information). In the non-twisted bilayer, the system is in a QSH state unless an extrinsic sufficiently large Semenoff mass term or an additional Rashba SOC term $\lambda_{\rm R}$ are added. In the 180$^\circ$ twisted case, the situation is very different regardless whether or not the intrinsic Semenoff mass term is compensated by an external electric field and regardless of $\lambda_{\rm SOC}$. Indeed, many changes in the SOC terms suppress the QSH state in the 180$^\circ$ twisted bilayer: The comparably large Rashba $\lambda_{\rm R}$ and the spin-valley coupling terms $B$ and a strong reduction in $\lambda_{\rm SOC}$. Each of these alone is sufficient to suppress the QSH state. At vertical electric field $E_{\rm z}=-4.2$~mV~\AA$^{-1}$ the gaps at K and K' close, and a band touching point emerges.

While it is clear that there will be tendencies towards interaction-induced (quasi)ordered phases as well, here, the kind of ordering is likely different from the non-twisted case but largely unexplored. The band touching at $E_{\rm z}=-4.2$~mV~\AA$^{-1}$ implements a situation similar to saddle points in a two-dimensional dispersion, where already arbitrarily weak interactions would trigger different kinds of magnetic or excitonic instabilities \cite{kotov_RMP_2012}. How these instabilities translate into the intermediately correlated and strongly spin-orbit coupled case of 180$^\circ$ twisted TaSe$_2$ is a completely open question.

\subsection{Conclusions and outlook}
The field of twistronics with materials like bilayer graphene is based on the idea that weak interlayer coupling can flatten highly dispersive bands and thereby boost electronic correlations \cite{bistritzer_macd_pnas_2011,cao_unconventional_2018,cao_correlated_2018}. The system introduced here takes the opposite route of deconfining formerly Mott localized electrons. This approach should be generally applicable to interfaces of Mott localized electrons under two conditions: the interlayer coupling should substantially exceed the in-plane one and at the same time define a connected graph linking all sites of the system. Possible example systems range from stacking faults in the bulk of CCDW layered Mott materials \cite{Kourkoutis_TEM_2016} to molecular systems \cite{tsukahara_evolution_2011}. 

Especially the twisting degree of freedom opens new directions to experiments. Since interlayer hopping is the dominant kinetic term in deconfined Mott systems like bilayers of CCDW 1T-TaSe$_2$, we expect incommensurability effects to be much more pronounced than in twisted graphene systems \cite{bistritzer_macd_pnas_2011}. $\theta=30^\circ$ twisted CCDW 1T-TaSe$_2$ bilayer should realize a quasicrystal with twelvefold rotation symmetry and provide an experimental route to correlated electrons and emerging collective states in a quasicrystalline environment.

The prototypical case of CCDW 1T-TaSe$_2$ bilayer demonstrates how deconfinement of Mott electrons leads to exotic states of quantum matter: the non-twisted bilayer approaches a quantum tricritical region of competing QSH, trivial band insulating, and antiferromagnetic insulating states. At 180$^\circ$ twist angle different kinds of electrically controllable band degeneracies with associated many-body instabilities, hypothetically of excitonic type, emerge. Clearly, the phase space for manipulating deconfined Mott electrons is high dimensional. We here identified the combination of twist angle and perpendicular electric field as decisive for TaSe$_2$ bilayers. Further means to control emerging electronic states include dielectric engineering \cite{pizarro_internal_2019} and charge doping. Our calculations showed that the non-twisted CCDW 1T-TaSe$_2$ bilayer in honeycomb stacking approximates the (Kane-Mele) Hubbard model on the honeycomb lattice with $U/|t|\approx 3.8$ very well. In this regime, doping the system away from the Dirac point towards the van Hove singularity is expected to lead to chiral superconductivity \cite{black-schaffer_resonating_2007,nandkishore_chiral_2012,kiesel_competing_2012,black-schaffer_chirald-wave_2014}, most likely of $d+id$-type.


\beginmethods
\textbf{DFT calculations.} We perform DFT \cite{PhysRev.136.B864,PhysRev.140.A1133} calculations by using the Vienna \textit{ab-initio} simulation package (VASP) \cite{Kresse_1994,PhysRevB.59.1758} with the generalized gradient approximation of Perdew, Burke, and Ernzerhof (GGA-PBE) for the exchange-correlation functional \cite{PhysRevLett.77.3865,PhysRevLett.78.1396}. We obtain the total energies, relaxed structures and electronic structure of mono- and bilayer 1T-TaSe$_2$ systems. We calculate the total energies for various possible stackings in undistorted and CCDW bilayer 1T-TaSe$_2$ using $\Gamma$-centered \textbf{k}-meshes of 15$\times$15$\times$1 and 9$\times$9$\times$1, respectively and taking into account van der Waals (vdW) corrections within DFT-D2 and cross-checking with DFT-D3 \cite{doi:10.1002/jcc.20495,grimmedftd3}, see Supplemental Figure 1. The ionic relaxation is done using the conjugate gradient algorithm until all force components are smaller than $0.02$~eV~\AA$^{-1}$. 

Since the DFT-D2 corrections yield correct interlayer distances but do not correctly capture the CCDW distortions, we adopted the following relaxation procedure to calculate the commensurate $\sqrt{13} \times \sqrt{13}$ CCDW bilayer structures:
\begin{enumerate}
\item We perform relaxations for a $\sqrt{13} \times \sqrt{13}$ supercell of the monolayer (without vdW corrections), using a superlattice constant of $a=12.63$~$\text{\AA}$ according to Ref. \cite{chen2019visualizing}. We fix the vertical positions of the Ta atoms, while allowing for Ta in-plane displacements. The Se atoms are allowed to freely relax in all three directions.

\item We include then a second layer and optimize the interlayer distance, $d$, while keeping all relative intralayer distances fixed. We find $d=6.4$ $\text{\AA}$ for the ideal honeycomb stacking in CCDW bilayer 1T-TaSe$_2$.

\item We relax the CCDW bilayer following the same procedure described for the monolayer, i.e. without vdW corrections and vertical positions of the Ta atoms according to the optimized interlayer distance $d$ fixed, while allowing for in-plane displacements. The Se atoms are allowed to freely relax in all three directions.
\end{enumerate}

We cross check the results obtained by this procedure against calculations with vdW corrections according to the DFT-D3 method. \cite{grimmedftd3}.
In contrast to DFT-D2, the CCDW distortions are well described in the DFT-D3. Thus, full relaxations of all atomic positions have been performed in the DFT-D3 framework for the mono- and bilayer. Both our step-by-step procedure using DFT-D2 method, and the full relaxation using DFT-D3 give equivalent results for the total energies (see Supplemental Figure 1b), for the crystal and band structures.

For the non-collinear magnetic calculation, i.e. when SOC is included, we set the net magnetic moment to zero in all atoms of the unit cell, and use a $\Gamma$-centered \textbf{k}-mesh of 6$\times$6$\times$1.

\textbf{Estimation of the screened Hubbard interaction $\mathbf{U}$ via RPA.} We estimate the local Hubbard interaction $U$ for the flat bands around the Fermi level in the CCDW 1T-TaSe$_2$ bilayer from the \textit{ab-initio} calculation of the screened Coulomb interaction, using RPA for the undistorted bilayer. We follow a similar procedure as in Ref. \cite{Kamil_2018}, which we summarize below:
\begin{itemize}
\item We initially calculate the \textsc{Wannier90} tight-binding model for the three low-energy Ta bands $\mathcal{C}$, whose orbital character is mostly $\{d_{z^2},d_{x^2-y^2},d_{xy}\}$ (see Fig. 2(a) in Ref. \cite{Kamil_2018}) in the undistorted monolayer 1T-TaSe$_2$. 
\item The static RPA-screened Coulomb interaction tensor $W_{\alpha \beta \gamma \delta}(\textbf{q},\omega \rightarrow 0)$ is calculated for undistorted monolayer 1T-TaSe$_2$, where $\textbf{q}$ is a reciprocal wave vector on a $\Gamma$-centered mesh of 18$\times$18$\times$1, and $\alpha,\beta,\gamma,\delta \in \mathcal{C}$. We neglect $\textbf{q}=\textbf{0}$ terms in our RPA analysis in order to avoid unphysical  effects.
\item In the CCDW 1T-TaSe$_2$ bilayer, the $d_{z^2}$ orbitals from Ta atoms in the SoD centers have the largest contribution for the bands around the Fermi level. Thus, for each $\textbf{q}$, we consider only the tensor element $W(\textbf{q}) \equiv W_{\alpha\alpha\alpha\alpha}(\textbf{q})$ with $\alpha=d_{z^2}$.
\item Then, the local Hubbard interaction $U$ in a single star of David is calculated by averaging over the $d_{z^2}$ orbital weight from each Ta atom (labeled by $w_{d_{z^2}}(\textbf{R})$) in the star of David:
\begin{equation}
U=\sum_{\bold{R},\bold{R}' \in \davidsstar} w_{d_{z^2}}(\textbf{R}) U(\textbf{R}-\textbf{R}') w_{d_{z^2}}(\textbf{R}')
\end{equation}
where $U(\textbf{R})$ is the discrete Fourier transform of $W(\textbf{q})$.
\end{itemize}

\textbf{DMFT and TPSC many-body calculations.} For non-twisted CCDW 1T-TaSe$_2$ bilayer, we construct tight-binding-Hubbard Hamiltonians of the type
\begin{equation}
	H_\mathrm{TBH} = \sum_{\langle i, j\rangle, \sigma,\sigma^\prime} t_{ij}^{\sigma\sigma^\prime} c_{i\sigma}^\dagger c_{j\sigma^\prime} + U \sum_i n_{i\uparrow} n_{i\downarrow}
\end{equation}
from a Wannierization of the \emph{ab-initio} DFT band structure (Supplementary Information), and estimate the on-site repulsion $U$ to be about 130\,meV  by means of RPA calculations. We study these effective low-energy models from a many-body perspective to judge the type of correlations in the system. In DMFT the lattice Hamiltonian is mapped onto a self-consistently determined single impurity problem, solved -- in our case -- within numerical exact quantum Monte Carlo in the hybridization expansion flavour (CT-HYB) (for a review, see \cite{Gull}). The resulting sublattice-resolved self-energy, $\Sigma$, is local (${\bf k}$-independent) but it contains frequency-dependent non-perturbative corrections beyond Hartree-Fock to all orders and can account for Mott-Hubbard metal-to-insulator transitions. All DMFT calculations are performed using w2dynamics \cite{w2dynamics}. The double counting is accounted for using the fully-localized limit. For two-dimensional systems it is important to estimate non-local effects at the level of the self-energy, not included in DMFT. To this goal, we apply the TPSC method \cite{Vilk_1997}, which produces accurate results in the weak-to-intermediate coupling regime, if compared to lattice quantum Monte Carlo calculations in the single band Hubbard model. For non-twisted CCDW 1T-TaSe$_2$ bilayer, which is modelled by a multi-band system we use the multi-site formulation of TPSC~\cite{Zantout2018} while neglecting the Hartree term to avoid double counting of correlation effects already accounted for in DFT. Moreover, to be able to apply TPSC to this system we project out spin off-diagonal terms and take only the diagonal spin-up contributions from DFT while still assuming a paramagnetic state. The combination of TPSC accounting for the ${\bf k}$-dependence of $\Sigma$ and DMFT, in which we can include all off-diagonal terms between spin-orbitals and access antiferromagnetic ordering at strong coupling consitutes a powerful tool to determine the many-body nature of 1T-TaSe$_{2}$.

\begindatav
The data that support the findings of this study is available from the corresponding author upon reasonable request.

\beginacknowd
We thank D. Di Sante, P. Eck, E. van Loon, M. Sch\"uler, and C. Steinke for useful conversations. JMP and TW acknowledge funding from DFG-RTG 2247 (QM$^3$) and the European Graphene Flagship. SA and GS are supported by DFG-SFB 1170 Tocotronics, and further acknowledges financial support from the DFG through the W\"urzburg-Dresden Cluster of Excellence on Complexity and Topology in Quantum Matter -- \textit{ct.qmat} (EXC 2147, project-id 390858490). TM, KZ and RV acknowledge funding from the DFG through grant VA117/15-1. PB acknowledges financial support from the Italian Ministry for Research and Education through PRIN-2017 project ``Tuning and understanding Quantum phases in 2D materials - Quantum 2D'' (IT-MIUR Grant No. 2017Z8TS5B). This research was supported in part by the National Science Foundation under Grant No. NSF PHY-1748958. We gratefully acknowledge the Gauss Centre for Supercomputing e.V. (www.gauss-centre.eu) for funding this project by providing computing time on the GCS Supercomputer SuperMUC at Leibniz Supercomputing Centre (www.lrz.de). Computing time at HLRN (Berlin and G\"ottingen) is acknowledged.

\begincompint
The authors declare no competing interests.

\beginauthorcont
J.M.P. performed the DFT and RPA calculations. S.A. calculated the topological invariant diagrams. S.A., K.Z. and T.M. performed the DMFT and TPSC calculations. P.B. derived the $k \cdot p$ model. R.V., G.S. and T.W. supervised the project. J.M.P., S.A., G.S. and T.W. analyzed the results. All authors contributed to write the paper.



\beginsupplement

\section{Stacking potential energy landscapes}
\label{suppl:sec:DFTWann}

\begin{figure*}
\includegraphics[width=0.9\linewidth]{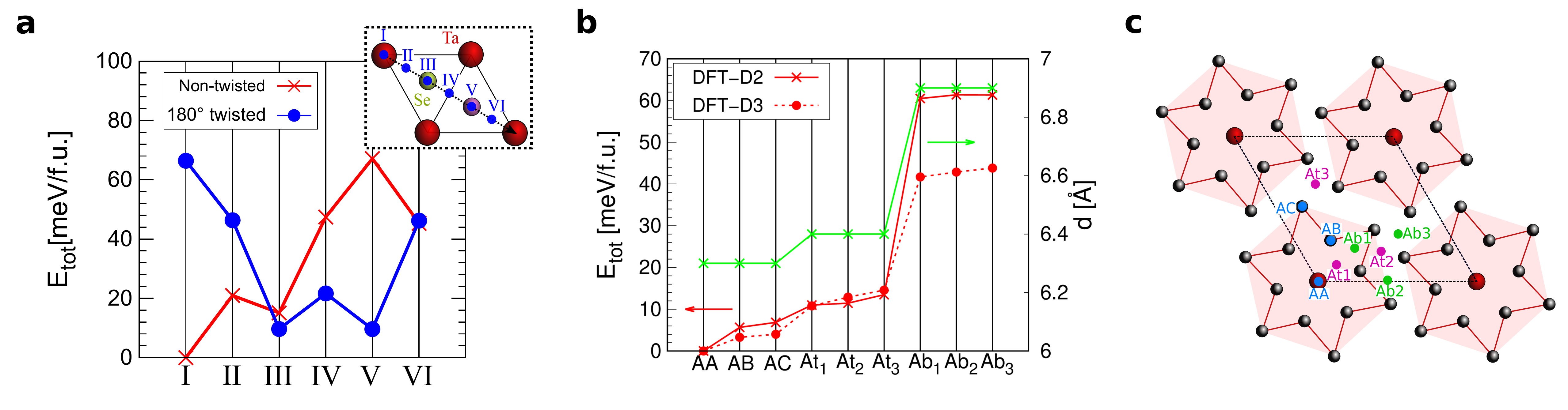}
\caption{\textbf{Stacking potential landscapes for 1T-TaSe$_2$ bilayers.} \textbf{a,} Potential energies for undistorted 1T-TaSe$_2$ bilayer in the non-twisted and 180$^{\circ}$ twisted cases. The inset shows the nomenclature of the different stacking configurations I-VI with bottom layer Ta atoms in red and Se atoms above and below the bottom Ta plane in green and pink, respectively. The stacking of the top layer is marked with blue dots and latin numbers. Blue dots refer to the position of the Ta atom in the top layer, starting from the perfectly aligned bilayer in I. Non-twisted bilayer shows a total minimum for stacking I (perfect alignement), and a local metastable minimum for stacking III (Ta on top of the Se above the Ta plane). For 180$^\circ$ twisted bilayer, stackings III and V are degenerate and are the most stable configurations. In the ideal honeycomb stacking, the CCDW 1T-TaSe$_2$ bilayers have a local stacking with the Ta in the top layer (approximately) above Se in the bottom layer, i.e. configurations III and V. \textbf{b,} Total energies per formula unit and interlayer distances $d$ versus stacking configurations in the CCDW non-twisted bilayer 1T-TaSe$_2$ as obtained within the DFT-D2 and DFT-D3 approaches. \textbf{c,} Nomenclature used for the different CCDW stackings, where blue, green and magenta dots describe the location of the Star-of-David (SoD) centers from the top layer. Thus, stacking AA corresponds to perfectly aligned CCDW bilayer. Stackings AtX stand for SoD central Ta atom from the top layer aligned to the Se atom above the Ta plane from the bottom layer, and AbX for SoD center from the top layer above the Se atom below the Ta plane from the bottom layer. DFT-D2 and DFT-D3 calculated energies are similar, with a discrepancy of approximately 20~meV per f.u. for the AbX stackings. For Ta on top of Ta (AA, AB, AC) and Ta on top of Se above the lower Ta plane (At1, At2, At3) energy differences are $\lesssim$10~meV per f.u.. Stacking At3 is the one considered in the main text.}
\label{figs1:potentialstack} 
\vspace{-0.3cm}
\end{figure*}

Here we study the dependence of the total energies of 1T-TaSe$_2$ bilayers on the local atomic stacking (Supplementary Figure \ref{figs1:potentialstack}a) and the stacking of the CCDW centers (Supplementary Figure \ref{figs1:potentialstack}b). 

Dependencies on the local atomic stacking can be inferred from the stacking potential energy landscapes derived from our DFT calculations for undistorted bilayers of 1T-TaSe$_2$. The non-twisted and 180$^{\circ}$ twisted cases are shown in Supplementary Figure \ref{figs1:potentialstack}a. For non-twisted bilayer, Ta from the top layer above Ta from the bottom layer (stacking I) is the most stable configuration. Top layer Ta above bottom layer Se (stacking III) results to be a local minimum. For 180$^{\circ}$ twisted bilayer, the most stable configurations are given for Ta on top of Se (stackings III and V). Regarding the local atomic arrangement, the honeycomb stacking discussed in the main text corresponds to configuration III.

Energy differences between different stacking configurations of the CCDW centers in the distorted non-twisted 1T-TaSe$_2$ bilayers are given in Supplementary Figure \ref{figs1:potentialstack}b with the labeling of the configurations being defined in Supplementary Figure \ref{figs1:potentialstack}c: AA, AB and AC refers to CCDW centers of the top layer on top of a Ta atom from the bottom layer (local configuration I). Both, AtX and AbX refer to arrangements, where the Ta atom in the CCDW centers from one layer is on top of / beneath Se atom sites in the other layer (local configurations III and V, respectively). In this notation, At3 is the honeycomb stacking that is considered in the main text.

It becomes clear that the undistorted bilayer (Supplementary Figure \ref{figs1:potentialstack}a) already gives a good overall estimate of the total stacking differences energies of the CCDW bilayer (Supplementary Figure \ref{figs1:potentialstack}b). 
Hence, also in the CCDW case, the stacking potential energy landscape is dominated by contributions from the local atomic stacking.


In the CCDW non-twisted bilayer, the total stacking energies using DFT-D2 and DFT-D3 calculations are similar, with a small deviation for the most unfavorable stackings AbX (local configuration V). The resulting crystal and band structures in both methods (not shown) are also equivalent. The energy differences between Ta on top of Ta (AA, AB, AC) and Ta on top of Se above the Ta plane (At1, At2, At3) are relatively small, i.e. on the order of 10~meV per f.u.. This means that the potential energy landscapes are very flat, which is in line with realizations of multiple stacking configurations in experiments.

Assuming that the 180$^{\circ}$ twisted bilayer is similar to non-twisted case in that the potential energy landscape is dominated by contributions from the local atomic stacking, we expect that stackings AtX and AbX are the most stable ones in the 180$^{\circ}$ twisted case. This would support formation of the honeycomb pattern in the 180$^{\circ}$ twisted case, while again flat potential energy landscapes are expected.

\section{Wannier tight-binding model}
\label{suppl:sec:Wannier}
The relevant subspace $\mathcal{B}$ for the low-energy bands of (distorted) CCDW 1T-TaSe$_2$ monolayer and bilayer contains mainly $d_{z^2}$ orbitals from Ta atoms in the centers of the stars-of-David (SoD, see Fig. 1). We construct a corresponding minimal tight-binding model using \textsc{Wannier90} code \cite{MOSTOFI2008685}. The Wannier basis is $\mathcal{B}=\{ d_{z^2}^{top},d_{z^2}^{bottom} \}$ for each spin, where the superscripts refer to SoD center Ta atom from top and bottom layers. We use a $\Gamma$-centered $\textbf{k}$-mesh of 9$\times$9$\times$1 if SOC is not included in the calculation, and $\Gamma$-centered $\textbf{k}$-mesh of 6$\times$6$\times$1 when SOC is included. In the latter case, an inner energy window covering states at K (and K') and M is considered, while leaving out the ones at $\Gamma$. We show in Supplementary Figure \ref{figs2:wannier} the comparison between the DFT bands and our Wannier tight-binding model. Our model captures very well the low-energy bands of non-twisted and twisted CCDW 1T-TaSe$_2$ bilayers.

\begin{figure*}
\includegraphics[width=0.9\linewidth]{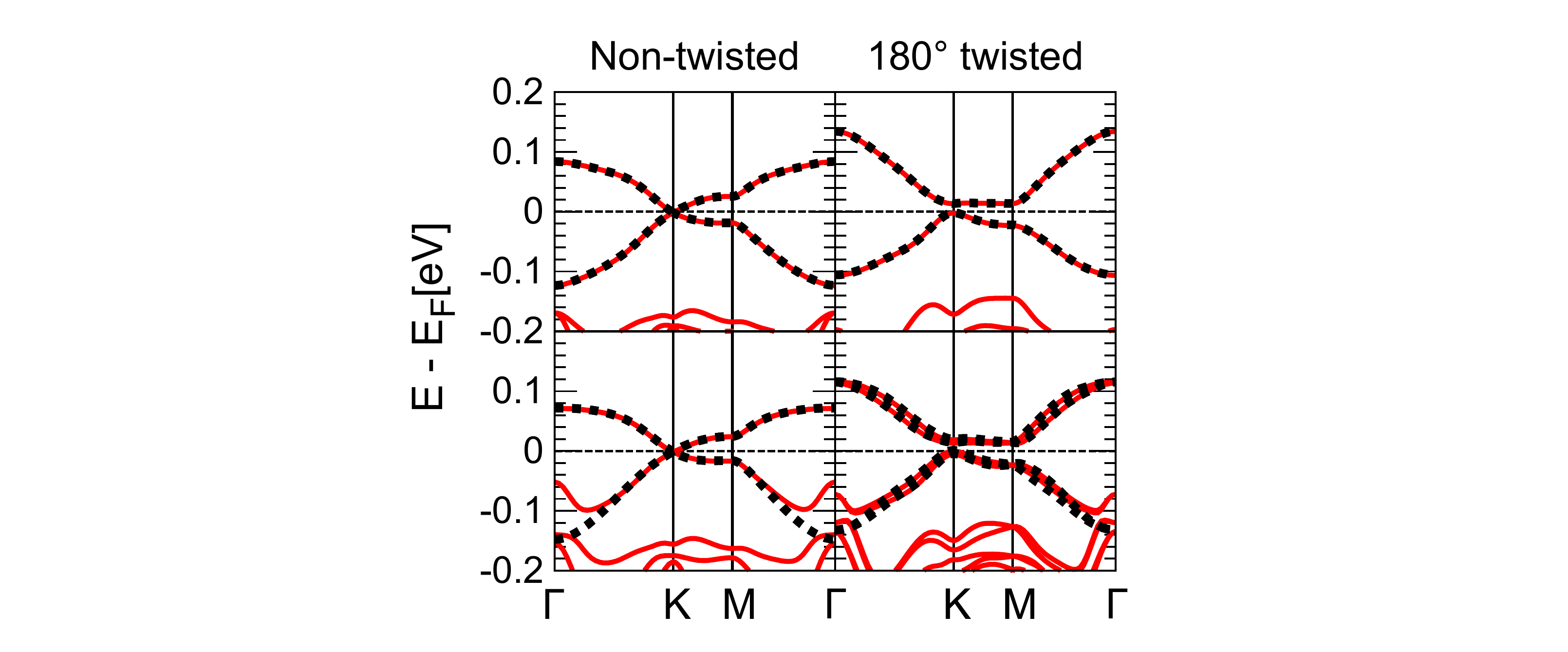}
\caption{\textbf{\textit{Ab-initio} DFT band structure versus Wannierization for non-twisted and 180$^{\circ}$ twisted CCDW 1T-TaSe$_2$ bilayer}. DFT bands (red solid lines) are compared with our Wannier tight-binding model (black dashed lines) for non-twisted (left panels) and 180$^\circ$ twisted (right panels) bilayers. Top panels show the case without SOC included, the bottom panels with SOC taken into account. Our Wannier tight-binding models fit the DFT band structures obtained for CCDW 1T-TaSe$_2$ bilayers very well.}
\label{figs2:wannier} 
\vspace{-0.3cm}
\end{figure*}

\section{$\pmb{k \cdot p}$ model of low energy bands}
\label{suppl:sec:k.p}
We derive the effective model describing the low-energy band structure around the K and K' points by using the $\mathbf{k} \cdot \mathbf{p}$ perturbation theory \cite{cardona,dresselhaus} for the perturbed Hamiltonian  $H=H_{\rm I}+H_{\rm{II}}+H_{\rm{III}}$, including Pauli-type spin-orbit interaction, where:
\begin{eqnarray}
H_{\rm I} &=&\frac{\hbar}{m}\mathbf{k}\cdot \mathbf{p}\\
H_{\rm{II}} &=&\frac{\hbar}{4m^2c^2}(\nabla U\times \mathbf{p})\cdot \bm{\sigma} \equiv\frac{\hbar}{4m^2c^2}\mathbf{A}\cdot\bm{\sigma}\\
H_{\rm{III}} &=&\frac{\hbar^2}{4m^2c^2}(\nabla U\times\mathbf{k})\cdot\bm{\sigma} \equiv \frac{\hbar^2}{4m^2c^2}(\mathbf{k}\times\bm{\sigma})\cdot\nabla U.
\end{eqnarray}
The non-zero matrix elements of such perturbed Hamiltonian can be determined from group theory and exploiting the transformation properties of the one-electron wave functions under the symmetry operations of the high-symmetry point K. The knowledge of the little group of K allows to define a set of linear equations \cite{weiss_pr1958}:
\begin{eqnarray}\label{eq:symmelements}
\langle u^{\mu}_{\nu}\vert O^{\alpha}_{\beta} \vert u^{i}_{j}\rangle &=& \frac{1}{h}\sum_R \sum_{\nu'\beta'j'} {}^\mu D^\ast_{\nu'\nu}(R) {}^\alpha D_{\beta'\beta}(R)\nonumber\\
&& \times{}^i D_{j'j}(R)\,\langle u^{\mu}_{\nu'}\vert O^{\alpha}_{\beta'} \vert u^{i}_{j'}\rangle
\end{eqnarray}
where $h$ is the order of the group, $u^\mu_\nu$ represents the $\nu$-th component of the basis function of a representation $\mu$ and $ O^{\alpha}_{\beta}$ is an operator that transforms like the $\beta$-th component of the basis function of a representation $\alpha$, while  ${}^i D_{j'j}(R)$ is the $(j'j)$ element in the matrix representative of the group element $R$ in the $i$-th representation. The non-twisted (180$^\circ$-twisted) CCDW bilayer belongs to the $P\bar{3}$ ($P3$) space group; both structures display a three-fold rotation $C_3$ around an axis perpendicular to the bilayer, while the non-twisted stacked bilayer is also centrosymmetric, the two layers being inversion partners. At point K only $C_3$ symmetry is preserved, and non-relativistic bands can be grouped in a one-dimensional single-valued irreducible representation (IR) $K_1$ and in two-fold degenerate $K_2K_3$ IR for the space group $P\bar{3}$; when the symmetry is lowered to $P3$, bands belonging to $K_2$ and $K_3$ are not degenerate anymore. The single-valued IRs, the corresponding characters and the basis functions are listed in Table \ref{tab_kp}. By introducing a general operator $\bm \pi$ with components
\begin{eqnarray}
\pi_1 = p_z, \quad \pi_2 = \frac{1}{\sqrt{2}}\left(p_x+i\,p_y \right), \quad \pi_3 = \frac{1}{\sqrt{2}}\left(p_x-i\,p_y \right),
\end{eqnarray}
each trasforming as $K_1, K_2$ and $K_3$, respectively, one can use equation (\ref{eq:symmelements}) to identify its symmetry-allowed non-zero expectation values on the basis functions $\phi_2,\phi_3$ spanning the $K_2K_3$ IR:
\begin{eqnarray}\label{eq:matrixelements}
\langle \phi_2\vert\pi_3\vert\phi_3\rangle,\quad \langle \phi_3\vert\pi_2\vert\phi_2\rangle, \quad\langle \phi_2\vert\pi_1\vert\phi_2\rangle, \quad\langle \phi_3\vert\pi_1\vert\phi_3\rangle.
\end{eqnarray}
In this basis, therefore, the following $\mathbf{k}\cdot \mathbf{p}$ model is found to fulfill the point-group symmetries of the wave vector at K:
\begin{eqnarray}\label{eq:kpmodel_full}
H&=&\left(\begin{array}{cc}
\lambda_{1}\sigma_z+\alpha_{1}(k_x\sigma_y-k_y\sigma_x) & -\hbar v_f k_++\lambda_{R}(i\sigma_x-\sigma_y)+\lambda_D k_+\sigma_z\\
-\hbar v_f k_- -\lambda_R(i\sigma_x+\sigma_y) + \lambda_{D}k_-\sigma_z & \lambda_{2}\sigma_z+\alpha_{2}(k_x\sigma_y-k_y\sigma_x)
\end{array}\right)\nonumber\\
\end{eqnarray}
where $k_\pm=k_x\pm ik_y$ and we introduced the following parametrization:
\begin{eqnarray}
v_f &=&\frac{1}{2m}\langle \phi_2\vert p_x-ip_y\vert\phi_3\rangle \nonumber\\
\lambda_{1} &=& \frac{\hbar}{4m^2c^2}\langle \phi_2\vert A_z\vert \phi_2\rangle\nonumber\\
\lambda_{2} &=& \frac{\hbar}{4m^2c^2}\langle \phi_3\vert A_z\vert \phi_3\rangle \nonumber\\
\lambda_{R} &=& - i \frac{\hbar}{4m^2c^2}\frac{1}{2}\langle \phi_2\vert A_x-iA_y\vert\phi_3\rangle\nonumber\\
\lambda_{D} &=& -i \frac{\hbar^2}{4m^2c^2}\frac{1}{2}\langle \phi_2\vert (\nabla U)_x-i(\nabla U)_y\vert\phi_3\rangle\nonumber\\
\alpha_1 &=&\frac{\hbar^2}{4m^2c^2}\langle \phi_2\vert (\nabla U)_z\vert \phi_2\rangle \nonumber\\
\alpha_2 &=&\frac{\hbar^2}{4m^2c^2}\langle \phi_3\vert (\nabla U)_z\vert \phi_3\rangle.
\end{eqnarray}
The effective model must also obey time-reversal symmetry $\Theta=\hat{T}K$, where $K$ is the complex conjugation and $\hat{T}=i\sigma_y \idop$. However, since K point is not time-reversal invariant, acting with the $\Theta$ operation will map the Hamiltonian relative to point K to the time-reversal partner K'.
Additionally, inversion symmetry also must be enforced for the non-twisted honeycomb structure, where the inversion operation $\hat{P}=\idop S_x$, swapping the basis functions, also maps the Hamiltonian equation (\ref{eq:kpmodel_full}) from point K to point K'. Here $\mathbf{S}$ represents the sublattice pseudospin, spanning the two-dimensional space defined by the basis functions $\phi_2,\phi_3$. 

Imposing both time-reversal and inversion invariance, one finds that $\lambda_{R}=0$, $\lambda_{D}=0$ and:
\begin{eqnarray}
\lambda_{1}=-\lambda_{2} \equiv -\lambda_{\rm SOC} &\quad\quad& \alpha_{1}=-\alpha_{2}\equiv\alpha_{R2}
\end{eqnarray}
recovering the effective $\mathbf{k}\cdot\mathbf{p}$ model of equation (1) in the main text, describing the low-energy band structure around K, K' points of non-twisted honeycomb CCDW bilayer. Imposing only time-reversal symmetry, as relevant for the twisted CCDW bilayer, implies that all terms appearing in equation (\ref{eq:kpmodel_full}) are allowed by symmetry. Furthermore, the lack of inversion symmetry (coinciding here with a sublattice-symmetry breaking) removes all degeneracies of the unperturbed non-relativistic bands, thus introducing an effective mass term $M$ that acts as a staggered potential. By introducing the following parametrization:
\begin{eqnarray}
B = \frac{1}{2}\left(\lambda_{1}+\lambda_{2}\right) &\quad\quad& \lambda_{SOC}=\frac{1}{2}\left(\lambda_{1}-\lambda_{2}\right)\nonumber\\
\alpha_{R1} = \frac{1}{2}\left(\alpha_{1}+\alpha_{2}\right) &\quad\quad& \alpha_{R2}=\frac{1}{2}\left(\alpha_{1}-\alpha_{2}\right)
\end{eqnarray}
we recover the low-energy model equation (2) in the main text, describing the band structure in the vicinity of the K, K' points.

We conclude this appendix noticing that the low-energy Hamiltonian equation (\ref{eq:kpmodel_full}) almost coincides with the one derived in Ref. \cite{Kochan2017} for a graphene-based system with $C_{3v}$ symmetry, but for the additional spin-momentum coupling term parametrized by $\lambda_D$ and allowed here by the lower $C_3$ symmetry. In fact, a minimal tight-binding model reproducing the low-energy band structure in the vicinity of K and K' points can be derived following the general scheme outlined in Ref. \cite{Kochan2017} and taking into account the reduced symmetry. 
Alongside the intrinsic SOC (next-nearest neighbor spin-conserving hopping, that is sublattice dependent due to lack of sublattice symmetry, parametrized here by $\lambda_1,\lambda_2$), the lack of any horizontal reflection allows for the so-called ``pseudospin inversion asymmetric'' SOC \cite{gmitra2013} (next-nearest neighbor spin-flipping hopping, also sublattice dependent, parametrized here by $\alpha_1,\alpha_2$) and, together with inversion-symmetry breaking, for the Rashba SOC \cite{kanemele.prl95.2005} (nearest-neighbor spin-flipping hopping, parametrized by $\lambda_R$).  The {\it no-go} arguments of Ref. \cite{Kochan2017} also implies that the lack of any vertical reflections in the $C_3$ structural symmetry allows for a purely imaginary nearest-neighbor spin-conserving SOC hopping, whose coupling constant can be parametrized by $\lambda_D$.

\begin{figure}%
\vspace{0.50 cm}
\includegraphics[width=0.9\columnwidth]{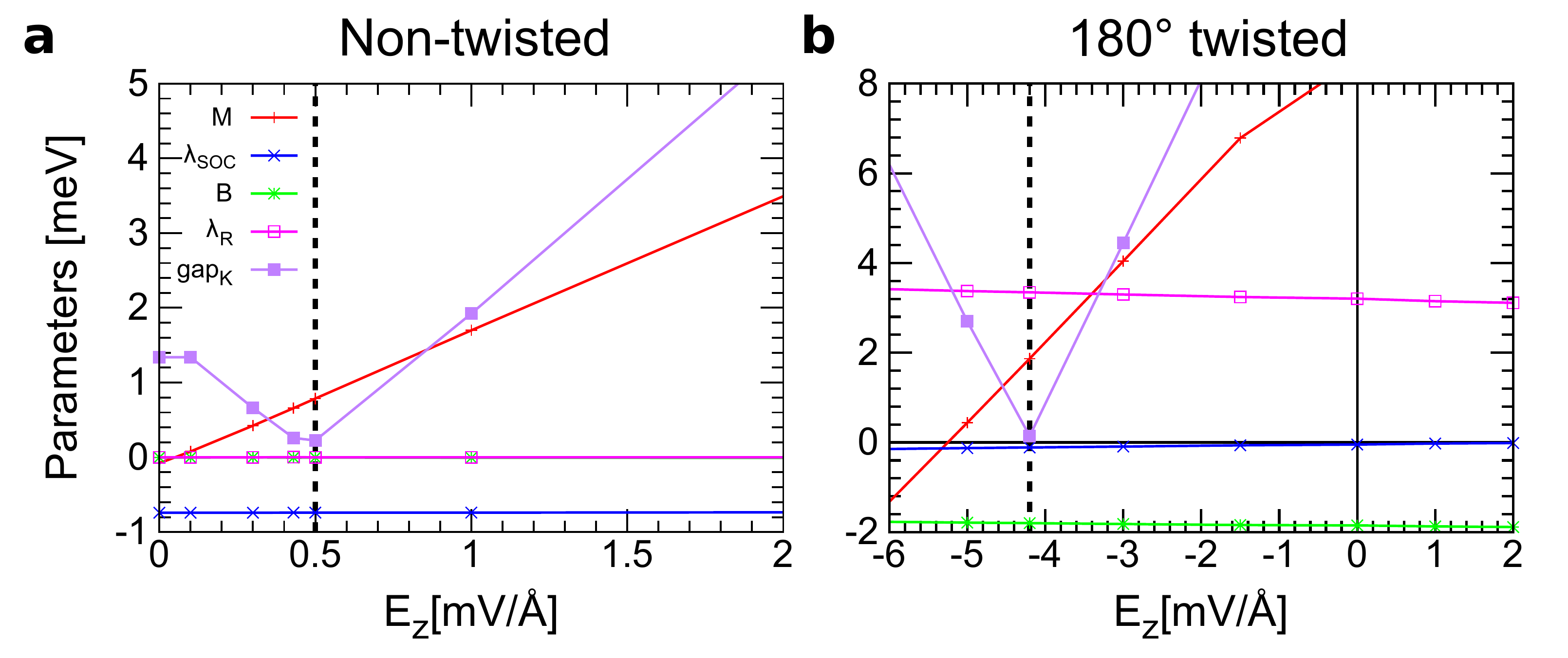}
\caption{\textbf{Parameters of the $\bm k \cdot \bm p$ expansion around Brillouin zone corners $K,K'$ of the Wannier Hamiltonitans}. Parameters for \textbf{a,} non-twisted and \textbf{b,} 180$^\circ$ twisted CCDW 1T-TaSe$_2$ bilayer in vertical electric fields. The Semenoff mass $M$, Kane-Mele SOC $\lambda_{\rm{SOC}}$, spin-valley coupling $B$, and Rashba SOC $\lambda_{\rm R}$ are shown as function of vertical electric field $E_{\rm z}$.}%
\label{fig:param_efield}%
\end{figure}

We obtain all effective parameters entering in the low-energy models around K and K’ from a first-order Taylor expansion in $k$-space of the Wannier tight-binding models described above in terms of the external field $E_z$, see Supplementary Figure \ref{fig:param_efield}.

\begin{table}[h]
\begin{tabular}{rl}
&
\begin{tabular}{c|p{1cm}|p{1cm}|p{1cm}|p{2.7cm}c}
&\centering $E$ & \centering $3^+$ & \centering$3^-$ & \centering basis functions&\\
\hline
$K_1$ &\centering 1 & \centering1 & \centering1 & \centering$P_z$, $A_z$&\\
\end{tabular}
\\
$\displaystyle{\biggl\{\Bigr.}$ &
\begin{tabular}{c|p{1cm}|p{1cm}|p{1cm}|c}
$K_2$ & \centering1 & \centering$\omega$ &\centering $\omega^\ast$& $P_x+i P_y$, $A_x+i A_y$\\
$K_3$ & \centering1 & \centering$\omega^\ast$ &\centering $\omega$& $P_x-i P_y$, $A_x-i A_y$
\end{tabular}
\end{tabular}
\caption{Character table for the little group at K. Here $\omega=e^{i\frac{2\pi}{3}}$, and $\bm P (\bm A)$ stands for polar (axial) vector. Notice that polar and axial vectors transform in the same way under the symmetry operations of $C_3$ point group.}\label{tab_kp}
\end{table} 

\section{Topology}
\label{suppl:sec:topology}

The topological properties of the twisted and non-twisted TaSe$_2$ bilayers without the effects of correlation are studied using the Wannier charge center evolution, as described in \cite{Soluyanov}. Using the Wannier tight-binding Hamiltonians as input, we determine the Z$_2$ invariant to assess whether they describe a trivial or a quantum spin Hall insulator. 

For the non-twisted structure we find that the  Z$_2$ invariant equals 1 for all values of the electric field between 0 and 0.43~mV~\AA$^{-1}$ \;while Z$_2$=0 for all other $E$-field strengths. For the 180$^\circ$ twisted bilayer we find Z$_2$=0 for all calculated $E$-field strengths instead.

In order to understand why the 180$^\circ$ twisted bilayer does not show any topologically non trivial phases we analyze the influence of the terms entering the Hamiltonian $H_{\rm 180^\circ}$ from equation (2) of the main text. We proceed from the Wannier Hamiltonian of the non-twisted bilayer and artificially add and vary parameters occurring in $H_{\rm 180^\circ}$ from equation (2). Specifically, we consider a parameter space, where we vary the Semenoff mass $M$, the Rashba-spin orbit term $\lambda_{\rm R}$ and the spin-valley coupling term / valley Zeeman $B$. To determine the topological properties of non-interacting Hamiltonians in this parameter space, it turns out to be numerically efficient to track the minimal gap in the band structure and to identify connected regions of the parameter space with strictly non-zero gap, which are bounded by a submanifold of the parameter space with zero gap. If one point inside has Z$_2 \neq 0$ (Z$_2= 0$) the whole region is topologically non-trivial (trivial). The resulting topological phase diagrams are shown in Fig. 3c of the main text. 

It is useful to compare our system to the well-known case of the ideal Kane-Mele model \cite{kanemele.prl95.2005}. The characteristic shape of the phase-diagram with the Rashba coupling ($\lambda_{\rm R}$) on the $y$-axis and the staggered potential $M$ (referred to as Semenoff mass) on the $x$-axis in the ideal Kane-Mele model resembles that of an onion \cite{kanemele.prl95.2005}. We observe this behavior for the non-twisted case, see leftmost panel of Fig. 3c. The yellow horizontal line represents the trajectory of our Hamiltonian upon changing the electric field $E_{\rm z}$. The main effect of $E_{\rm z}$ is to change $M$ according to $M\approx M_{\rm 0}+e E_{\rm z} d/\epsilon_{\rm \perp}$ (see Supplementary Figure \ref{fig:param_efield}) and hence $E_{\rm z}$ drives the 0$\circ$ twisted system from inside the topological region to the outside.

\begin{figure}
\includegraphics[width=0.45\textwidth]{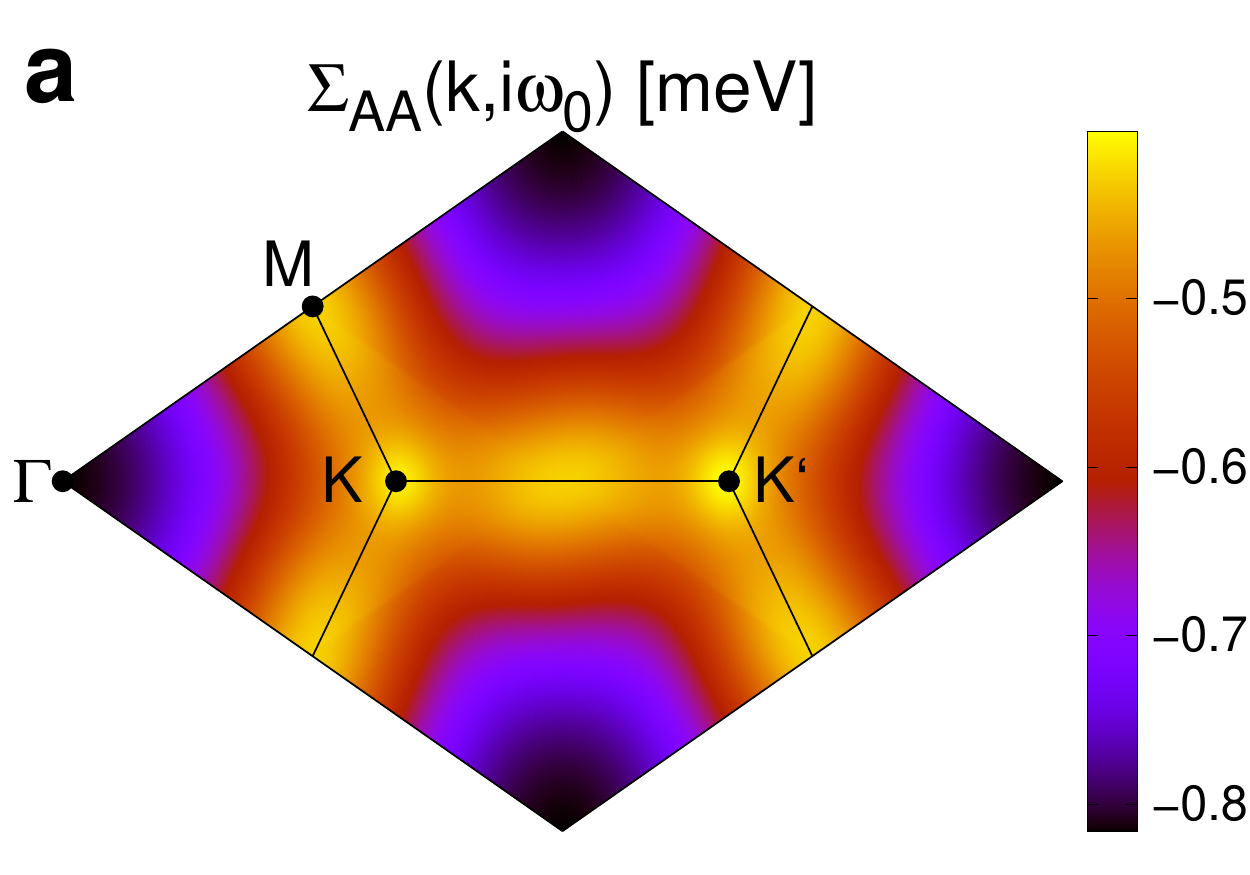}
\includegraphics[width=0.45\textwidth]{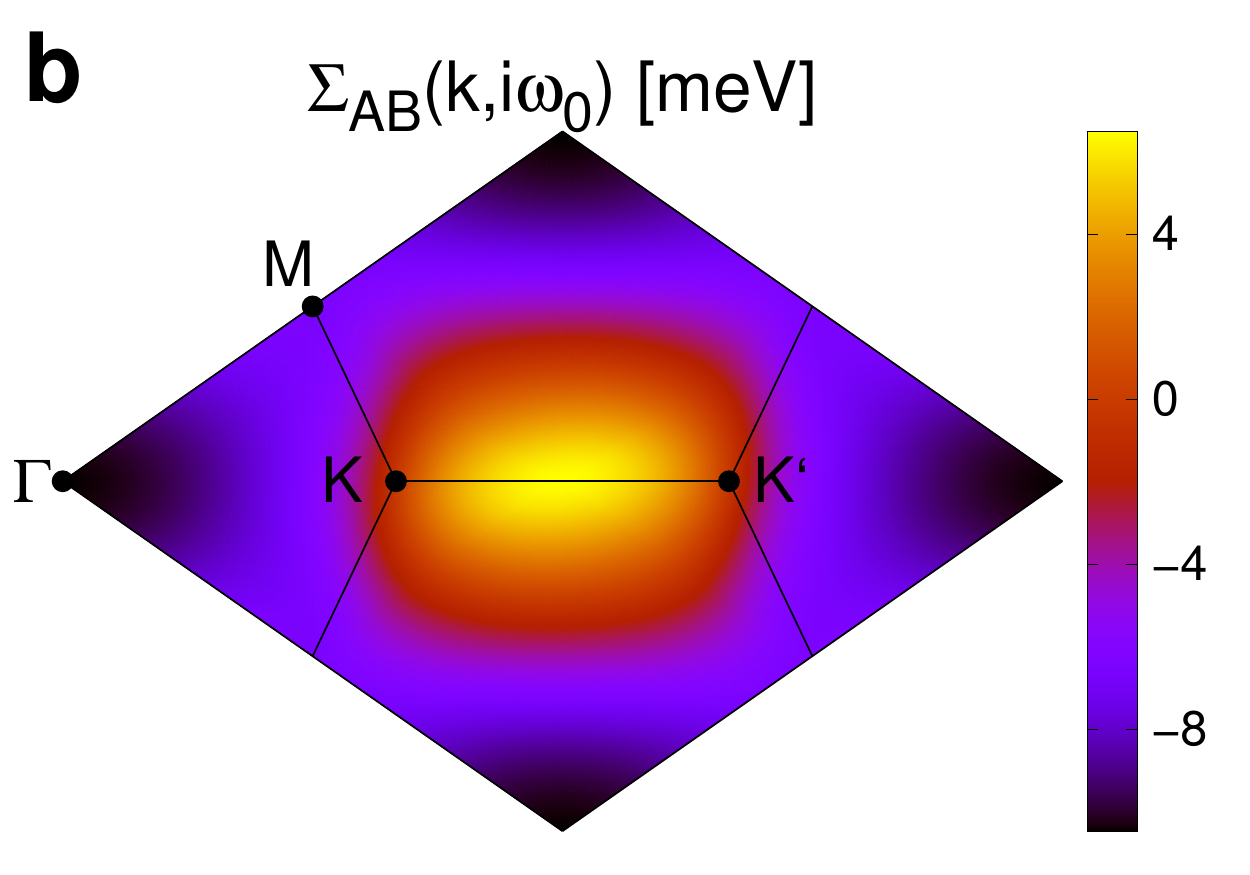}
\caption{\label{dmftVStpscZ}
\textbf{Momentum dependent intra- and inter-sublattice parts of the self-energy as obtained from the TPSC calculations}. The \textbf{a,} intra- and \textbf{b,} inter-sublattice parts of the self-energy are obtained at $T=0.005$~eV$ = 58$~K. The antiferromagnetic fluctuations lead to a significant non-local inter-sublattice self-energy $\Sigma_{\rm AB}$}.
\end{figure}

Upon twisting, we reduce the symmetry from $D_{6h}$ to $C_{3v}$ consequently switching on various terms in $H_{\rm 180^\circ}$, as described in Ref.~\cite{Kochan2017}, among which the most important ones are the previously-mentioned Semenoff mass $M$, Rashba-spin orbit $\lambda_{\rm R}$ and the spin-valley coupling $B$. The main effect of $B$ is to shift two of the bands having the same sublattice but different spin character at the K-point towards each other. This yields a deformed onion as topological region, where the topological region on the $\lambda_{\rm R}$ axis is reduced. (See Fig. 3c of the main text for the case of $B/\lambda_{\rm SOC}=0.9$.) Eventually when $|B|=|\lambda_{\rm{SOC}}|$ the topological non-trivial region is completely suppressed and only a vertical band touching line remains, which, however, does not separate a trivial from a non-trivial region. The latter case is similar to the phase diagram of the 180$^\circ$ twisted TaSe$_2$ bilayer. The effects of $B$ and $\lambda_{\rm R}$ can, thus, qualitatively explain the differences observed for the topological phases of the non-twisted and 180$^\circ$ twisted structure. In the 180$^\circ$ twisted structure, $\lambda_{\rm SOC}$ is additionally strongly reduced, which further contributes to suppressing the QSH phase. Following this line of argumentation the electric field in the 180$^\circ$ twisted case induces a gap closing and a reopening but does not induce a topological phase transition.



In the non-twisted case, where we find QSH states in absence of interactions, we study the impact of correlations on the topological phase diagram within the TPSC approach. These calculations confirm the generic shape of the schematic phase diagram shown in (Fig. 2e). 

It is known \cite{Griogio_first-order_2015} that sufficiently strong Coulomb repulsion $U$ can change the nature of the topological phase transition between QSH and band insulator from continuous with a band-touching point at the transition  -- as in the non-interacting Kane-Mele or Bernevig-Hughes-Zhang models -- to first-order with discontinuous jump of the gap (solid line in Fig. 2e). Since the electric field $E_{\rm z}$ directly controls $M$, one can possibly tune the non-twisted system through this exotic first-order transition in experiments by varying $E_{\rm z}$. 


\section{DMFT and TPSC}
\label{suppl:sec:dmft_tpsc}

As shown in the main text (Fig. 2), TPSC and DMFT consistently yield quasiparticle weights $Z\approx 0.75$ for temperatures in the range 60-230~K for non-twisted CCDW 1T-TaSe$_2$ bilayer. Also double occupancies $\langle n_{\downarrow} n_{\uparrow}\rangle\approx 0.15$ agree well between TPSC and DMFT in this temperature range. These results consistently put non-twisted CCDW 1T-TaSe$_2$ bilayer at intermediate local correlation strength and clearly far away from a paramagnetic Mott Hubbard transition.

The onset of non-local correlations can be inferred from the temperature dependent enhancement of antiferromagnetic susceptibilities and correlation lengths, which we obtained with TPSC and show in the main text (Fig. 2). These correlations manifest in sublattice off-diagonal contributions to the self-energy shown in Supplementary Figure \ref{dmftVStpscZ}.

\bibliography{CDWTaSe2_npjQM_2}
\bibliographystyle{natphys}


\end{document}